\documentclass[prl,twocolumn,aps, nofootinbib, keywords]{revtex4} 

\usepackage{amsmath}
\usepackage{graphicx}
\usepackage{dcolumn}
\usepackage{bm}
\usepackage{amssymb}
\usepackage{epstopdf}
\usepackage{xcolor}
\usepackage{comment}

\usepackage{xcolor}
\usepackage[colorlinks = true,
            linkcolor = blue,
            urlcolor  = blue,
            citecolor = blue,
            anchorcolor = blue]{hyperref}

\begin{document}
\def\P{\mathbf{P}}
\def\Q{\mathbf{Q}}
\newcommand{\ket}[1]{\left| #1 \right>} 

\title{A dialog between cell adhesion and topology at the core of morphogenesis}

\author{Adri\'an Aguirre-Tamaral$^{1,\dagger}$, Elisa Floris$^{1,\dagger}$ and  Bernat Corominas-Murtra$^{1}$}

\thanks{bernat.corominas-murtra@uni-graz.at}
\affiliation{
$^1$Department of Biology, University of Graz, Universit\"atsplatz 2, 8010 Graz, Austria
$^*$Author for correspondence$\\$
$^\dagger$Equal contribution
}

\keywords{cell adhesion, topology, rigidity, morphogenesis}

\begin{abstract}
During the development of an organism, cells must coordinate and organize to generate the correct shape, structure, and spatial patterns of tissues and organs, a process known as morphogenesis. The morphogenesis of embryonic tissues is supported by multiple processes that induce the precise physical deformations required for tissues to ultimately form organs with complex geometries. Among the most active players shaping the morphogenetic path are fine-tuned changes in cell adhesion. We review here recent advances showing that changes on cell adhesion, a local, pair-wise property defined at the cell–cell contact level has important global consequences for embryonic tissue topology, being determinant in defining both the geometric and material properties of early embryo tissues.
\end{abstract}
\maketitle

%
One of the most fundamental constraints underlying the process of morphogenesis is that the acquisition of non-trivial shapes by the forming tissues has to take place without an external agent inducing such changes \cite{smith1985developmental, alberch1989logic, Wolpert:2015}. In consequence, 
the fine-tuned deformations of the embryonic tissue 
must be the outcome of the collective and coordinated behaviour of different groups of cells which, while changing their mechanical properties in a consistent way, will eventually differentiate and give functionality to the emerging tissue \cite{Graner:1993_1, 
gumbiner1996cell, lecuit2007cell, oates2009quantitative, Staple:2010,clark2014stresses, graner2017forms, tsai2022adhesion}. In such processes, underlying core physical properties define the phenotypical frame \cite{bi2015density, Kim:2021, atia2021cell, Petridou:2021} over which selective pressures will act. 
At the structural level, one must take into account geometrical considerations related to tissue morphology and, even deeper in the identification of the rawest structure, the {\em topological} properties of the embryonic tissue \cite{Giammona:2021, Petridou:2021, fischer2023tissues, Fabreges:2024}. Roughly speaking, the identification of topological properties is a fundamental way to characterize the kind of structure we are dealing with: Whether the structure contains holes, whether the object is tubular-like or spherical-like or, more fundamentally, what the pattern of connection of the different core elements of the system looks like \cite{
Petridou:2021, fischer2023tissues}. Importantly, the topological characterization is performed regardless of the geometric details, although the topological structure acts as a constraint for the potential geometries. Being topology at such fundamental level of characterization, natural key questions arise: How are the different topological patterns regulated? Is there a feedback between topological structure and material properties? How are different topological phenotypes achieved in front of other potential ones?  Without neglecting the role of other mechanisms, in this short review we will try answer the above questions focusing specifically on the role of cell adhesion as a key morphogenetic driver triggering changes on the topology of the tissue organization, changes that will eventually lead to the emergence of functional phenotypic traits.
%
%

\begin{figure*}
\begin{center}
    \includegraphics[width=0.65\linewidth]{Fig1.pdf}
    \caption{Different packings of spheres, that could represent cells in contact, with their geometry (top) and their associated topologies (bottom). For the sake of simplicity, we only show connected configurations ---i.e., with no isolated spheres. With 2 spheres, only one configuration is possible. With 3 spheres, 2 topological configurations are possible, the spheres aligned or arranged in a triangle. Note that the first and the second packings of 3 spheres have {\em different geometries} ---in the second one a sphere is twisted upwards--- but {\em the same topology}, because the structure of contacts is the same ---a line. With 4 spheres we have 5 different connected topologies in 3D. If the packing topology changes along a given process, 
we say that there has been a {\em topological transition}. Colors are used to show the correspondence between cells and nodes of the network.
}
    \label{fig:Fig1}
\end{center}
\vspace{-7mm}
\end{figure*}
%
%
{\bf The topology of the network of cell-cell contacts}
One of the most fundamental characterizations of the structure of the embryonic tissue arises from the network of cell-cell contacts \cite{
escudero2011epithelial, 
Petridou:2021, fischer2023tissues, Fabreges:2024}. We refer to the {\em topology} of cell-cell contacts to a particular network structure defined from just considering cells and contacts among them, regardless of any geometric property at any scale. The structure of contacts, when one deals with $n\geq 3$ cells, may have several realizations, since there are several ways to arrange the cells ---see Fig. \ref{fig:Fig1}. Every of such realizations, to which we will refer to as $g_1, . . .,g_N$, will define a topology of interactions ---i.e., a network--- whose particular structure is grasped by the {\em adjacency matrix}, $A(g_k)$ \cite{newman2018networks}.  If we are only interested in the raw topological structure, the elements  $a_{ij}$ of the adjacency matrix will be just either one or zero, depending on whether the contact between cells $c_i,c_j$ exists or not, respectively, while if we are interested in additional aspects, the adjacency matrix will be proportional to the surface of contact.
Topological changes, also known as {\em topological transitions} ---see Box 1---, between structural cell configurations will be represented by changes in the adjacency matrix. 
%

\begin{figure}[ht!]
\begin{center}
\includegraphics[width=7.8cm]{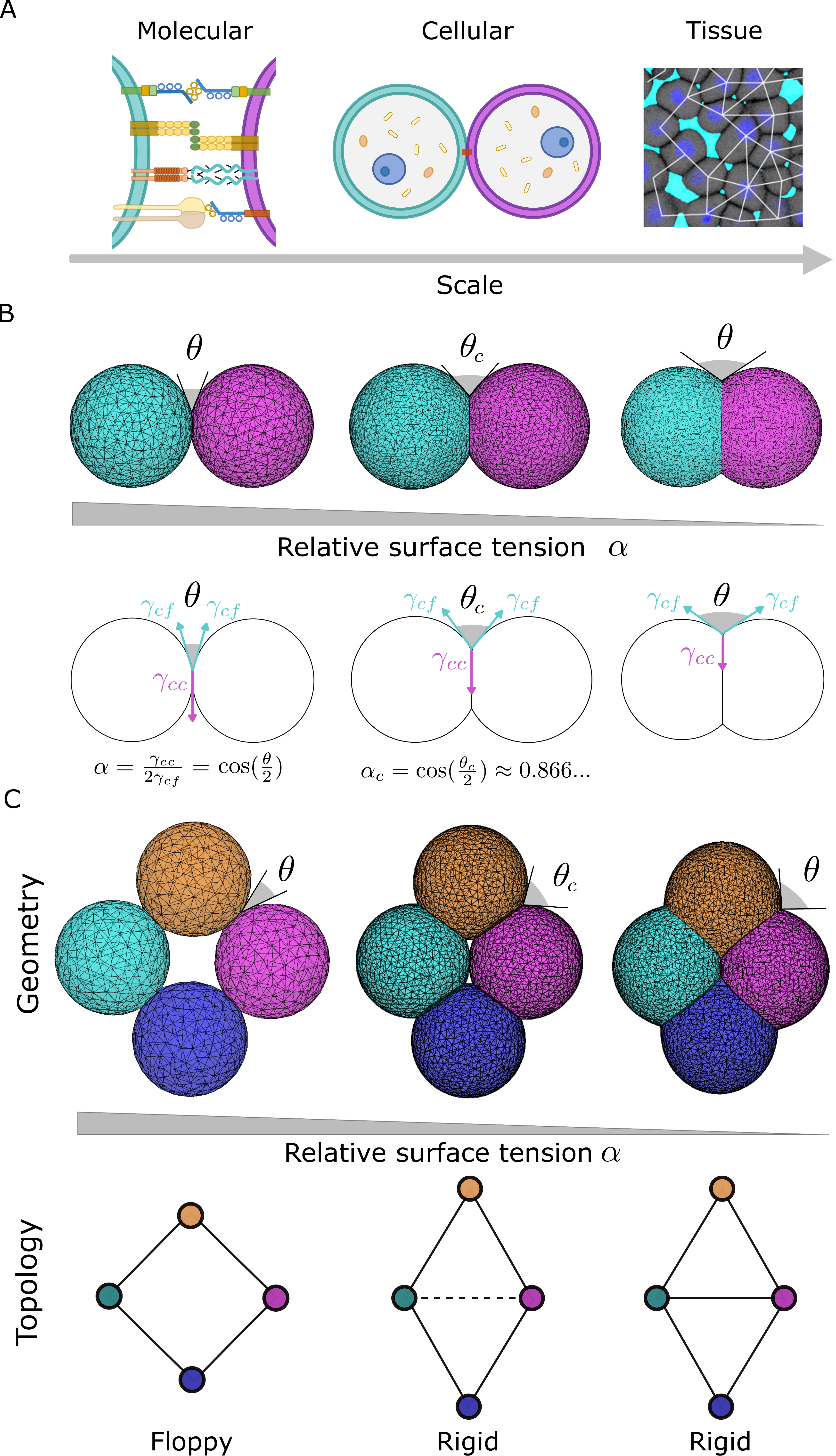}	
 \caption{Cell adhesion between cells. A) View of cell adhesion at different scales of description: micro (membrane level), meso (cell--cell level), and macro (tissue level with connectivity map).
B) Top: Changes in cell geometry induced by cell adhesion, characterized by the relative surface tension $\alpha$. Bottom: Schematic representation of how the cell-cell $\gamma_{cc}$ surface tension (pink arrow) and the cell-fluid $\gamma_{cf}$ surface tension (cyan arrows) vary as the relative surface tension $\alpha$ decreases, defined by the Young-Dupré relation, for two identical cells in contact. In force-balance conditions, $\alpha$ is empirically inferred from the contact angle between two cells $\theta$. C) Topological transition in which cells form a new connection at a critical value due to a reduction in surface tension $\alpha$. Top: geometrical representation of the cell configuration and associated changes in intercellular angle $\theta$. Bottom: Topological representation of the floopy-rigid
transition, in which a new link is created (dashed line). Colors are used
to show the correspondence between cells and nodes of the network.
}
\label{Fig:Fig2}	
\end{center}%
\vspace{-7mm}
\end{figure}

{\bf Cell adhesion}
The role of molecular cell adhesion is well known in biology as 
a fundamental mechanism for maintaining and shaping tissue architecture. Far from acting as a simple mechanical glue, cell adhesion works as a highly regulated mechanochemical process that coordinates tissue structure and maintenance \cite{gumbiner1996cell, halbleib2006cadherins, lecuit2007cell, clark2014stresses, Winklbauer:2015, tsai2022adhesion}.
Across metazoans, a substantial fraction of the genome encodes proteins associated with cell–cell and cell–matrix adhesion \cite{hynes2000evolution}. 
These cell adhesion molecules (CAMs) are typically grouped into several major families, including cadherins, integrins, immunoglobulin superfamily CAMs, and selectins. Generally, CAMs are transmembrane proteins that mediate adhesion outside the cell and connect to the cytoskeleton inside the cell \cite{gumbiner1996cell, joseph1998cell, hynes2000evolution}.
Through these connections, adhesion complexes couple chemical signals external to the cell, mechanical forces, and tissue-scale tensions to intracellular responses \cite{ salvi2018mechanisms, de2022interplay, Campas:2024}. 
In spite of the enormous complexity of the adhesion machinery described above, several mechanical phenomena can be studied from a simplified version of the membrane structure ---see Fig. \ref{Fig:Fig2}A. 

Indeed, 
for more than a decade, a wide range of abstract descriptions have been proposed. The widely used framework of vertex models, despite its simplicity, has been shown to capture several key properties of cell tissues \cite{Farhadifar:2007,Staple:2010, bi2015density,bi2016motility,damavandi2025universality}. To incorporate the presence of intercellular gaps in biological tissues, other computational approaches have been developed, including models of deformable particles \cite{boromand2018jamming,chiang2024intercellular}, self-propelled particles \cite{huang2023bridging}, 
cellular Potts models \cite{graner1992simulation, durand2019thermally}, and foam-like models. Here, we will focus on the adhesion dynamics studied 
 from the perspective of the theory of foams, establishing an analogy between soap bubbles with potentially variable surface tension properties and cells \cite{Weaire:1999, Hayashi:2004, Maitre:2012, Winklbauer:2015, Maitre:2015, parent2017mechanics, 
Kim:2021, ichbiah2023embryo, Fabreges:2024, firmin2024mechanics, Rustarazo:2025, vangheel2026rigidity}. In this framework, the key mechanical variable  describing two or more cells in contact with each other and with the interstitial fluid (non-confluent tissue)
is the relative strength of the cell-fluid  surface tension $\gamma_{cf}$ against the cell-cell $\gamma_{cc}$ surface tension, a non-dimensional parameter, $\alpha$, defined as:
\begin{equation}
\alpha=\gamma_{cc}/(2\gamma_{cf})\quad.
\label{eq:alpha}
\end{equation}
In force-balance conditions the parameter $\alpha$ can be inferred empirically from the contact angle between the two (identical) cells $\theta$ and the fluid as:
\begin{equation}
\alpha=\cos\left(\theta/2\right)\quad.
\end{equation}
The above relation is known as the {\em Young-Dupr\'e} relation, and it provides an empirically testable observable connecting theory and experiment ---see Fig. \ref{Fig:Fig2}B.
To explore the problem considering an arbitrary number of cells, one can define a rescaled Hamiltonian function for $c_1, . . ,c_n$ cells arranged in a specific configuration $g_k$ and a given relative surface tension between cells $\alpha$ as \cite{Weaire:1999,Brakke:1992,Maitre:2012}:
\begin{equation}
E_\alpha(g_k)=\alpha\sum_{i<j\leq n}a_{ij}+\frac{1}{2}\sum_{i\leq n}s_i+K\sum_{i\leq n}\left( v_i-v_0\right)^2 \quad,
\label{eq:hamiltonian_n}
\end{equation}
where $a_{ij}$ are elements of the adjacency matrix, corresponding to the contact surface between cells $i$ and $j$, $s_i$ stands for the area of cell $i$ in contact with fluid and the last term accounts for the (in)compressibility of the cells with volume $v$ around a preferred volume $v_0$, with $K$ accounting for the penalization of the volume deviations. The  works reviewed here assume nearly perfect volume conservation, i.e.~the constant $K$ is taken sufficiently large such that one can consider conservation of volume as a boundary condition.
 It is worth emphasizing that the parameter $\alpha$ intrinsically grasps the contributions of adhesion and contractility. Depending on the context, a decrease in relative surface tension $\alpha$ can result from a decrease in $\gamma_{cc}$ , potentially modelling an increase in cell adhesion, as schematically shown in Fig. \ref{Fig:Fig2}B, or from an increase in $\gamma_{cf}$ , potentially modelling an increase in contractility.  In these regards, and mainly due to the fact that tissues may not be confluent ---i.e., we allow gaps with interstitial fluid within the tissue---, adhesion is conceived in a slightly different way than in vertex models: In the latter, according to the energy function associated to them, increase of adhesion triggers an increasing of the contact surfaces, thereby increasing the perimeter/area relation  \cite{bi2016motility}. In consequence, in vertex models, increase of adhesion is interpreted as creating the energetic conditions that allow deformation of the membranes eventually enabling cell rearrangements at low energy costs.   
\vspace{2mm}

\noindent\fbox{%
    \parbox{0.46\textwidth}{
    \begin{center}
{\bf BOX 1: Statistical physics of topological transitions}
    \end{center}
    \footnotesize{ 
    Consider the set of different potential topologies $g_1, . . .,g_N$ made of $n$ cells and contacts therein, $\Omega(n)$, each with its energy in equilibrium $E_\alpha(g_i)$ given a relative adhesion strength $\alpha$. The probability of observing a certain configuration $g_i$ in equilibrium is:
    \begin{equation} 
    p_\alpha(g_i)=\frac{e^{-\beta G_\alpha(g_i)}}{\sum\limits_{g_i\in \Omega(n)} e^{-\beta G_\alpha(g_i)}}\quad,
    \label{eq:equilibrium}
    \end{equation}
    where $\beta$ acts as the inverse of the strength of random fluctuations in the contacts, i.e., plays the role of the inverse of an {\em effective} temperature ---not to be confused with the {\em thermodynamic} temperature, which is derived from stochastic fluctuations at smaller scales of the system. $G_\alpha(g_i)$ plays the role of free energy of the packing $g_i$. $G_\alpha(g_i)$ is derived from the contributions of the packing energy,  $E_\alpha(g_i)$, see eq. (\ref{eq:hamiltonian_n}) plus an additional term accounting for the configurational entropy, $S(g_i)$, which, roughly speaking, accounts for the existing symmetries within the packing \cite{meng2010free}, leading to:    
    \begin{equation}
    G_\alpha(g_i)=E_\alpha(g_i)-\frac{1}{\beta}S(g_i)\quad. 
    \label{eq:Gibbs}
    \end{equation}
 Under the assumption of detailed balance \cite{gardiner2009stochastic}, one can then compute the probability of transitioning between two configurations $g_i,g_j$ at constant $\alpha$:
    \begin{equation}
    \frac{p_\alpha(g_i\rightarrow g_j)}{p_\alpha(g_i\leftarrow g_j)}=e^{-\beta\left(G_\alpha(g_i)-G_\alpha(g_j)\right)}\quad.
    \nonumber
    \end{equation}
    Under changes of $\alpha$ the equilibrium distribution described in eq. (\ref{eq:equilibrium}) will change, and so will do the probabilities of transition from one state to another.    
    In consequence, upon changes on $\alpha$, some configurations that were highly probable may become less but, in turn, some other configurations may become more probable. If this happens, topological rearrangements will occur.
    
    }
    }
}
%

For $n$ cells some topologies may achieve lower or higher values of $E_\alpha$ and, crucially, the energetically favored configurations may change upon changes on $\alpha$.
In consequence, topological transitions, ---see Fig. \ref{Fig:Fig2}C--- in which e.g., new contacts are formed or old contacts are lost, are expected to occur. 
Again, we point out that the above described topological transitions may differ ---although not necessarily--- from the so-called T1 transitions occurring in vertex models \cite{Farhadifar:2007,Staple:2010, david2014tissue, bi2015density, damavandi2025universality}. In T1 transitions, a cell redefines its contact neighborhood, under the assumption of tissue confluency. In the topological transitions studied here, we do not assume confluency of the tissue by default  and, moreover, some of the properties that will be discussed can only be understood in the case of non-confluent tissues.
	
%

{\bf Topological rearrangement of early embryo packings}
We focus our attention on a case where topological transitions have been reported to be particularly relevant: {\em compaction} \cite{bowman1970cleavage, Wolpert:2015, gilbert2023developmental}.
Embryo compaction occurs at different developmental points in mammals, generally around the first 3--6 division rounds after fertilization. Along this process, not only the overall geometry of the blastocyst suffers deep changes, but also the topological structure of cell contacts.  The relevance of the topology to understand the biophysics of blastomeres has been highlighted since decades ago \cite{hillman1972effect, graham1979formation,  Giammona:2021, kuang2022computable, skinner2023topological, ichbiah2023embryo, Fabreges:2024}. 

In the 16-cell stage, a difference between outer vs. inner cell populations is established. Exterior cells will later form the trophectoderm, i.e., the extra-embryonic tissues, like the placenta, and inner cells will further develop into the organism itself \cite{Wolpert:2015}.  In consequence, compaction prepares the system at the 8-cell stage such that, in the next (stochastic) division round, a correct inner-outer cell proportion will be achieved. However, this raises questions like: What is the {\em right} configuration at the 8-cell stage and how can one identify it? Once this structure is identified, how do embryos undergo such a morphological convergence? 
$\\$
$\\$
\begin{figure*}
\begin{center}
\includegraphics[width=15cm]{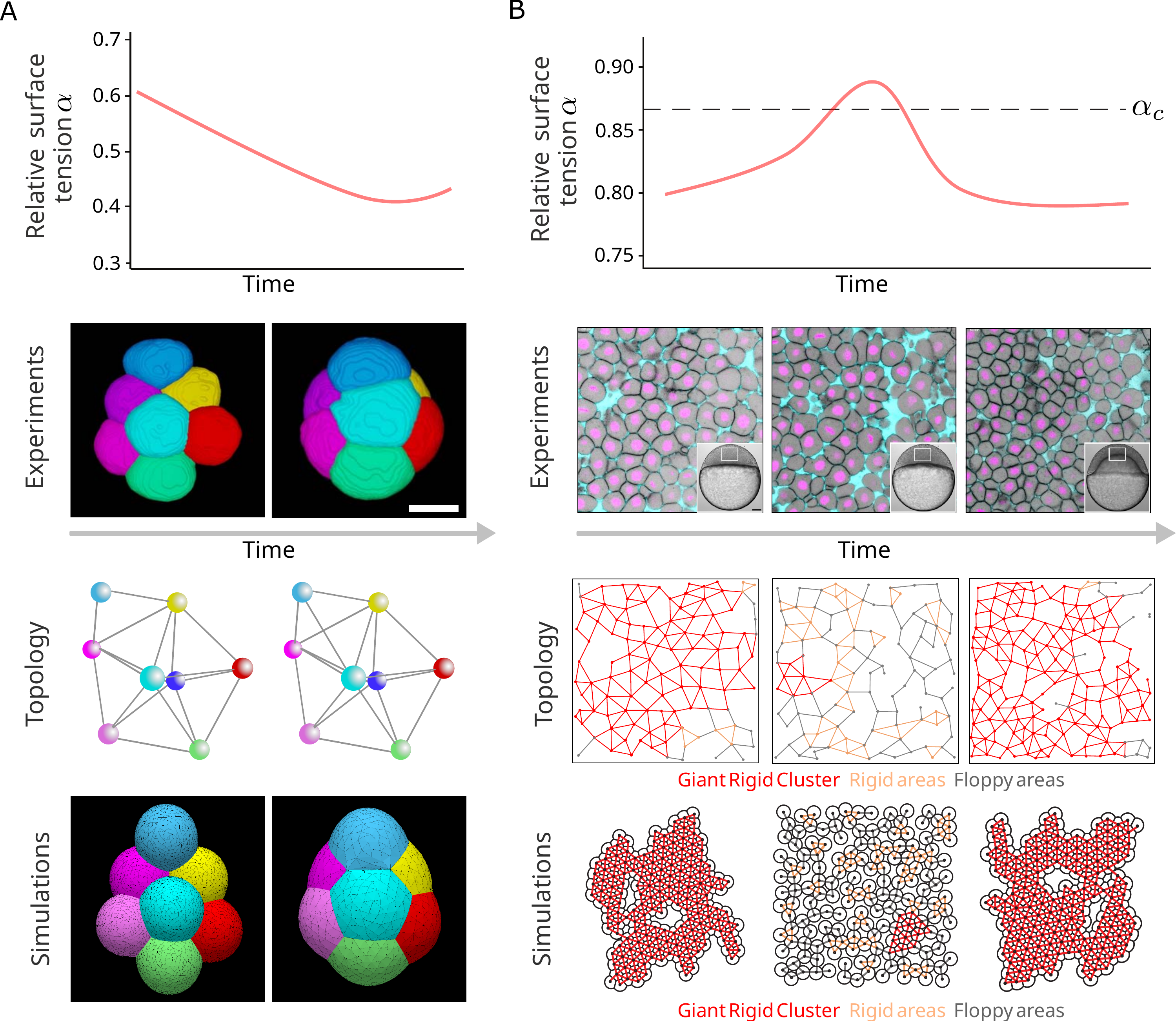}		
 \caption{ Biological implications of topological transitions.
A) Temporal evolution of the relative surface tension $\alpha$ during early mouse embryo development (plot adapted from \cite{Fabreges:2024}). Changes in $\alpha$ have a measurable impact on cell compaction, as observed in experimental 3D confocal reconstructions of the embryo, and at the topological level through the formation of new links in the cell contact network. This topological transition can be understood and simulated (bottom) by taking into account the experimentally observed changes in cell adhesion $\alpha$ (top). Imaging data of real embryos from T. Hiiragi's lab (Scale bar, 25 $\mu m$).
B) Temporal evolution of the relative surface tension $\alpha$ during zebrafish embryo development (plot adapted from \cite{Petridou:2021, Rustarazo:2025}) (Scale bar full embryo inset, 100 $\mu m$). Variations in $\alpha$ lead to measurable changes in tissue configuration, as observed experimentally using 2D confocal images, and tissue properties, as reflected in network rigidity, where the GRC (in red) can either span the entire tissue or cover only a small fraction of it. These topological rigid-floppy configurations can be understood and simulated (bottom) by taking into account the experimentally observed values in cell adhesion $\alpha$ (top) \cite{Rustarazo:2025}. Imaging data of real embryo tissues from N. Petridou's lab.
}
\label{Fig:Fig3}	
\end{center}	
\vspace{-7mm}%
\end{figure*}  
\begin{figure*}
\begin{center}
\includegraphics[width=14.5cm]{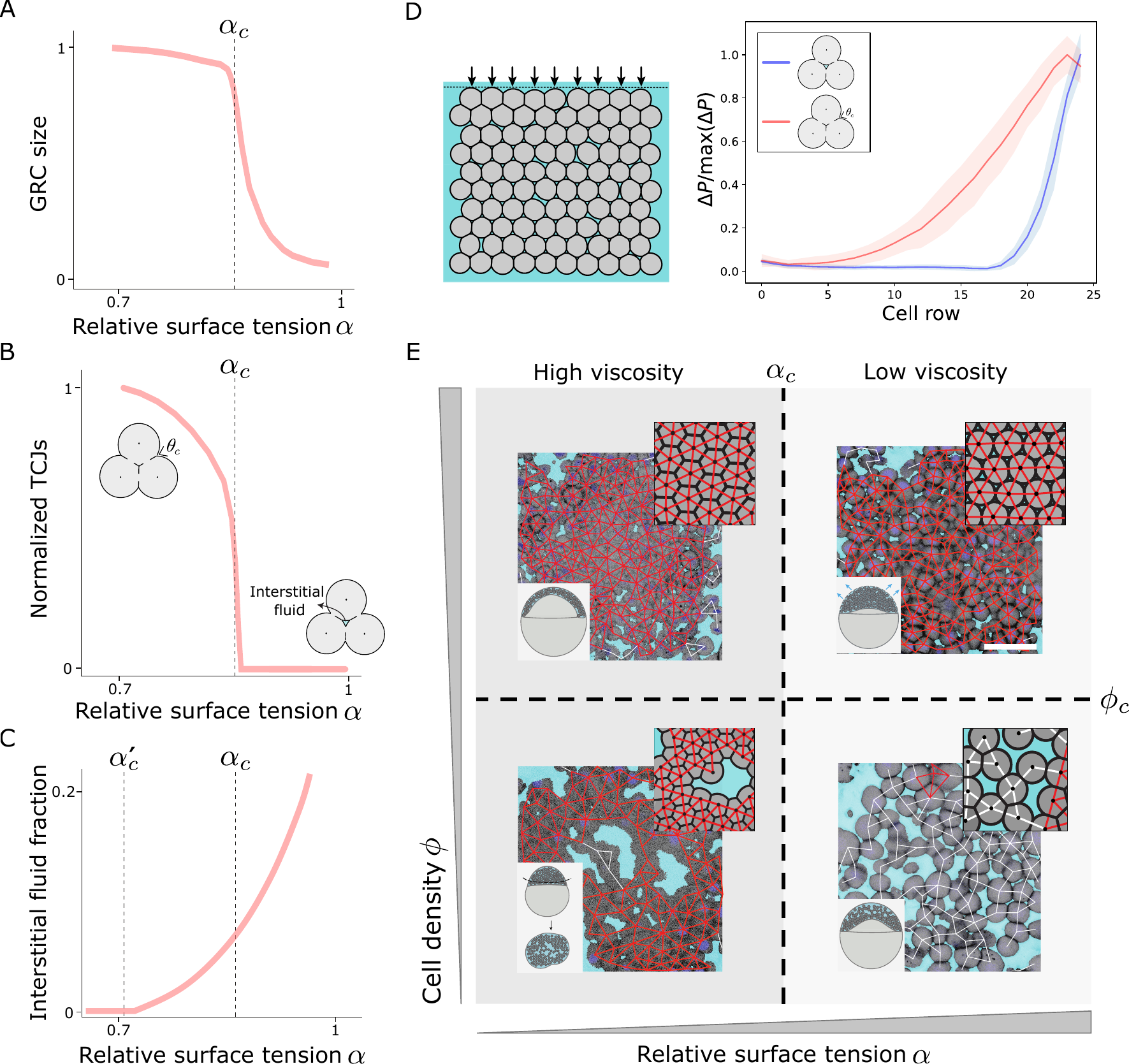}		
 \caption{Tissue properties driven by cell adhesion.
A) Phase transition showing the emergence of the GRC as a function of the relative surface tension $\alpha$ (plot adapted from \cite{Rustarazo:2025}).
B) Formation of TCJs in the tissue, defined as the closure of triangular configurations without interstitial fluid, which also undergoes a transition as a function of $\alpha$, interestingly occurring at the same critical value $\alpha_c$ triggering the emergence of the GRC (plot adapted from \cite{Rustarazo:2025}).
C) Quantification of interstitial fluid in tissues in force balance conditions as a function of $\alpha$, revealing two critical points, associated with the closure of triangular configurations ($\alpha_c$, TCJ formation) and to the closure of residual 3-dimensional octahedral fluid pockets ($\alpha'_c$, 3D pocket closure) (plot adapted from \cite{Autorino:2025}).
D) \texttt{Surface evolver} simulations showing the tissue response to applied pressure, comparing configurations 
where gaps between three cells are open to those where the gaps are closed and TCJs have formed.  $\Delta P$ is the pressure difference between the internal pressure of cells after and before compression. Network size: $25\times 25$, average statistics over 20 replicas, $\alpha=0.92$ for tissue simulation with open gaps between three cells 
(blue) and $\alpha=0.7$ for tissue simulation with 
formed TCJs (red), shaded region standard deviation.
E) Phase diagram of the tissue states as a function of cell density and relative surface tension. Experimental and simulated tissue configurations are shown across the different regions defined by the critical values ($\phi_c$, $\alpha_c$). The cell-cell contact network is shown in white and the GRC is highlighted in red. 
Imaging data of real embryo tissues from N. Petridou's lab, figure adapted from \cite{Rustarazo:2025} (Scale bar, 50 $\mu m$).}
\label{Fig:Fig4}	
\end{center}			
\vspace{-4mm}
\end{figure*}
In mouse, at the beginning of the 8-cell stage, a high degree of topological heterogeneity is observed. Then, as long as the compaction process unfolds, all the cell packings become topologically {\em rigid} \cite{pollaczek1927gliederung, Laman:1970, jacobs1995generic} ---See Box 2 for terminology and concepts on rigidity--- and, thus, the astronomic amount of topological configurations possible with 8 spheres are drastically reduced to 13 \cite{Arkus:2009, arkus2011deriving} ---see Box 2. Thanks to the Sch\"oenflies classification for the characterization of chrystallographic structures \cite{dorset1995modern}, we can identify a converging topology at the end of the compaction phase as the D2d, followed in abundance by the Cs(2) topology \cite{Arkus:2009, arkus2011deriving}. Interestingly, embryos moved through a fairly predictable sequence of topological transitions, reducing topological variability over time \cite{Fabreges:2024}. 

During compaction, the relative surface tension $\alpha$, also referred in mammalian early embryo development as {\em compaction} parameter, decreases \cite{Maitre:2012, Maitre:2015,ichbiah2023embryo,Fabreges:2024} ---see Fig. \ref{Fig:Fig3}A. This relative increase on cell adhesion is triggered actually by a sharp increase on the cell-fluid surface tension ---$\gamma_{cf}$ in eq. (\ref{eq:alpha})---, mediated by the actomyosin protein cortex \cite{Maitre:2015}. 
This, in turn, alters the energy balances and, under the frame provided by eq. (\ref{eq:hamiltonian_n}), triggering topological transitions towards more energetically favourable packings ---see Fig. \ref{Fig:Fig3}A. 
Under increasing adhesion, as observed in real embryos, the D2d topology becomes the lowest-energy configuration, and most initial packings converge to D2d in simulations, often via an intermediate visit to Cs(2) topology, as also observed in reality. To provide a counterfactual validation, it was shown that embryos that lack the maternal allele Myh9, required to generate surface tensions, failed to converge to a specific topology  \cite{Fabreges:2024}. Interestingly, stochastic variability at the initial stages of the 8-cell period was shown to facilitate the convergence to the D2d structures.

%
{\bf Rigidity transitions at tissue scale regulated by cell adhesion changes}
The effect of changes on the (microscopic) mechanical properties of the cells has been explored paying particular attention on how they change the topology of the tissue and, how this feeds back, in return, into the (macroscopic) material properties of the tissue \cite{Kim:2021, Petridou:2021, Rustarazo:2025}. For tissues with a larger number of cells, the amount of configurations becomes quickly astronomical. However, thanks precisely to the large amount of potential structures, a new approach to the study of the system is possible, considering the potential, expected properties of an {\em ensemble} of tissues sharing some structural features that could be i.e., the average number of contacts per cell.  
This statistical physics-like perspective over the topological properties of tissues has been considered in studying the onset of morphogenesis, right after the cleavage period in Zebrafish development \cite{Petridou:2021, Rustarazo:2025}. In this precise moment, known as {\em Doming phase}, the embryonic tissue switches, in a few minutes, from a high viscosity ($\eta\approx 400$ Pa$\cdot$s) to a low viscosity ($\eta\approx 25$ Pa$\cdot$s) state \cite{petridou2019fluidization,Petridou:2021}. This enables a smooth deformation of the tissue, covering the yolk cell \cite{kimmel1995stages}. 
The sudden drop in viscosity \cite{petridou2019fluidization} resembled a {\em phase transition} \cite{stanley1971phase, Petridou:2021}. Analysis of the network structure of cell-cell contacts revealed that, at the fluidization point, the average connectivity dropped slightly below the critical point of generic rigidity ---see Box 2---, showing that the framework of rigidity percolation \cite{jacobs1995generic} could account for the observed behaviour ---see Box 2. 
Using the relation between geometry and mechanical properties provided by the Young Dupré relation ---see eq. (\ref{eq:alpha})---, as well as large scale tissue simulations, one can establish a critical point in adhesion, $\alpha_c=0.866...$, at which a massive amount of links is formed, triggering a {\em rigidity percolation} phase transition ---see Box 2--- thereby rigidifying the tissue \cite{Petridou:2021, Rustarazo:2025} ---see Fig. \ref{Fig:Fig2}C and Fig. \ref{Fig:Fig3}B. In consequence, a slight drop/increase on the cell-cell adhesion is projected, at the tissue level properties, in a highly non-linear way ---see Fig. \ref{Fig:Fig4}A.

Analysis of cell density, understood as the fraction of space occupied by cells vs. interstitial fluid, revealed that changes in $\alpha$ potentially inducing rigidification (fluidization) matched also jamming (unjamming) values in cell fraction $\phi$ \cite{bolton1990rigidity, 
o2002random, van2010jamming}  ---see Box 2. In consequence, one could argue that, in Wild Type conditions, crossing the critical point in adhesion actually implies crossing the jamming critical point in cell density, $\phi_c\approx 0.84$, and that the actual driver of the rigidification/fluidization processes in non-confluent tissues is related to their cell density properties. However, in some conditions these two control parameters can be decoupled. In fact, it is possible  to engineer \textit{in vivo} embryos whose density lays way beyond the jamming threshold $\phi> \phi_c$ but display low viscosity \cite{Rustarazo:2025} ---a phenomenon observed in tumoral tissues \cite{ilina2020cell}. Conversely,  it has been also shown that it is possible to engineer \textit{in vivo} embryos whose tissue density lays below the critical fraction, $\phi<\phi_c$ but, still, display high viscosity \cite{Rustarazo:2025}. In consequence, the relative surface tension $\alpha$ reveals as the key driver of the tissue material properties, as, within the scale of values reported above, no tissue with $\alpha>\alpha_c$ displays high viscosity, and no tissue with $\alpha<\alpha_c$ displays low viscosity, regardless of the density $\phi$.

Under these 
observations, the effects of crossing $\alpha_c$ have been explored in detail. Aside of the massive creation of cell-cell contacts, a crucial change occurring when crossing $\alpha_c$ towards low values is the formation of the three-cell-junctions (TCJs) 
\cite{parent2017mechanics, Kim:2021, Rustarazo:2025, Autorino:2025}.
Thus, above $\alpha_c$ and in force balance conditions we expect, in any triangular configuration of cells,  to observe a small pocket of fluid whereas, at values below $\alpha_c$, such pocket is expected to be closed and a TCJ is formed, presumably introducing new structural constraints over the system ---see Fig. \ref{Fig:Fig4}B.
Still, if the 3D nature of the tissue is considered, pockets of fluid may remain there, which should be totally expelled at a second critical point, $\alpha_c'=0.71...$, in which a 3D tissue in force balance condition should be totally confluent \cite{Autorino:2025} ---see Fig. \ref{Fig:Fig4}C.

Therefore, empirical and theoretical findings point to the hypothesis that TCJs
, triggered at $\alpha_c$ , have an active role in defining the structure and material response of the tissue. To exemplify that, a numerical experiment of tissue compression using \texttt{Surface evolver software} \cite{Brakke:1992} shows that, in tissues with the same a-priori cell fraction $\phi>\phi_c$, stress propagation, expressed in terms of relative increase of internal pressure in cells, displays a much larger scale in tissues below $\alpha_c$, in which TCJs are formed, than in ones above $\alpha_c$, the ones containing empty spaces in the central part of triangular structures ---see Fig \ref{Fig:Fig4}D. 

As a conclusion, a {\em phase diagram} has been proposed, using the coordinates $\alpha,\phi$. This phase diagram contains four differentiated regions articulated around a double critical point in adhesion and density $(\alpha_c,\phi_c)$ \cite{Rustarazo:2025} ---see Fig. \ref{Fig:Fig4}E. These four regions, in turn, may refer to primitive versions of different tissue architectures: 1) $\alpha<\alpha_c$, $\phi>\phi_c$ {\em epithelial-like}, 2) $\alpha<\alpha_c$, $\phi<\phi_c$ {\em lumen-like}, 3) $\alpha>\alpha_c$, $\phi>\phi_c$ 
{\em mesenchymal-like} (dense), and 4) $\alpha>\alpha_c$, $\phi<\phi_c$ {\em mesenchymal-like} (sparse) \cite{Rustarazo:2025}. In summary, this phase diagram grasps both structural and material properties, and defines a discretized space of possibilities on how these properties may look like in real tissues.


{\bf Conclusions and outlook}
The theoretical analysis based on tissue topology provides abstract tools to bridge the microscopic level of cell-cell interactions and the emerging macroscopic tissue structure and properties. A central question is, in consequence, how developing organisms regulate tissue topology and, with it, its phenotypic outcomes. This report focuses on cell adhesion, highlighting recent studies which focused on how smooth and continuous changes at cell–cell interfaces propagate nonlinearly at the tissue scale, promoting convergence toward specific phenotypes and, in larger systems, defining a discrete set of possible tissue architectures and material properties. Further research efforts should try to connect the topological/mechanical properties of the tissue with the dynamics of cell fate decision \cite{Autorino:2025} or pathological scenarios \cite{ilina2020cell,hannezo2022rigidity}, thereby linking the emergence of function and morphogenesis.

\vspace{3mm}
\noindent\fbox{%
    \parbox{0.47\textwidth}{
    \begin{center}
{\bf BOX 2: Topological rigidity, basic concepts}
    \end{center}
    \footnotesize{

      -  Generic (topological) rigidity: A network structure made of nodes and connecting links is considered to be generically rigid if no independent (geometric) movement of the nodes is possible without stretching/compressing a link \cite{jacobs1995generic}. 
      $\\$
      
      - Connection between generic rigidity and viscosity: Interpreting links as viscoelastic springs, a generically rigid network will have Young modulus larger than zero, meaning that the structure will react against any deformation. For non-generically rigid networks, deformations are possible at no energy cost. In consequence, generically rigid tissues would be expected to display high viscosity, whereas non-generically rigid (floppy) ones would be expected to display a low viscosity or fluid-like response  \cite{Corominas:2021}.
                $\\$

        - Conditions for rigidity: The conditions under which a topological structure is generically rigid depend on the geometrical dimension the structure is embedded in \cite{pollaczek1927gliederung,Laman:1970, arkus2011deriving} ---see Fig. \ref{Fig:Fig2}C. In 2 dimensions we have that, for a network of $n$ nodes: 1) all nodes must be connected to at least $2$ other nodes and 2) there must be, at least $l$ links, in which $l$:
        \[
        l_{2D}\geq 2n-3\quad.
        \]
        In the case of structures embedded in 3 dimensions, the requirements are that 1) all nodes are connected to at least 3 other nodes and 2) the amount of links $l$ satisfies that:
        \[
        l_{3D}\geq 3n-6\quad.      
        \]        

        - Rigidity percolation: It is a high order {\em phase transition}. Once the average number of contacts per cell reaches a critical value, a generically rigid region that spans almost the whole network, the {\em Giant Rigid Cluster} (GRC), suddenly emerges. A small GRC implies that the structure is floppy, and a large GRC implies that the structure, as a whole, is rigid. In 2 dimensions, the critical point in the average number of contacts per cell triggering the emergence of the GRC is located at $\langle k_c\rangle=4$ for large networks \cite{jacobs1995generic}.
        $\\$
        
        - Jamming: A system ---made e.g., of hard disks--- is jammed when the constituents arrest each other and no independent movements are possible \cite{liu1998jamming, bolton1990rigidity}. At the jamming critical point in density $\phi$ of hard disks, the network of contacts experiences a sudden emergence of a large GRC ---see above. The critical point in density that leads to jamming of hard disks is $\phi_c\approx 0.84$, meaning that, in a tiling made of disks spread at random over a 2D plane, if the area covered by disks $\phi$ is such that $\phi\geq \phi_c$, the system will be jammed.
    }
    }
}

\newpage
{\bf Acknowledgments}
We thank Dimitri F\'abreges, Takashi Hiiragi and Nicoletta I. Petridou for their comments and generously sharing imaging pictures of \textit{in vivo} embyonic tissues. We also thank Artemy Kolchinsky for his constructive comments on the manuscript. This work was supported by the Weave project "Tissue material phase
transitions and their role in embryo pattern formation" from the Deutsche
Forschungsgemeinschaft (DFG, German Research Foundation, 518354236, PE 3800/1-1)  and  Österreichischer Wissenschaftsfonds (FWF, Austrian Science Fund, I6533) (E. F and B. C-M). B.C-M. and A. A-T. acknowledge the support of the field of excellence “Complexity of
life in basic research and innovation” of the University of Graz.

\bibliographystyle{unsrt}


\end{document}